\documentclass[12pt]{iopart}
\usepackage{color}
\usepackage{epsfig}
\usepackage{iopams}
\usepackage{epsfig}
\usepackage{graphicx}
\usepackage{graphics}
\usepackage{verbatim}
\newcommand{\eq}{\begin{eqnarray}}
\newcommand{\en}{\end{eqnarray}}

\usepackage{color}

\begin{document}

\vspace*{-2cm}
\title{Masses and widths of scalar-isoscalar multi-channel resonances
from data analysis}

\author{Yurii S. Surovtsev$^1$,
        Petr~Byd\v{z}ovsk\'y$^2$,
        Robert~Kami\'nski$^3$,
        Valery E. Lyubovitskij$^{4,5}$
        Miroslav~Nagy$^6$}

\address{
$^1$ Bogoliubov Laboratory of Theoretical Physics, Joint
Institute for Nuclear Research, 141 980 Dubna, Russia\\
$^2$ Nuclear Physics Institute, Czech Academy of Sciences,
\v{R}e\v{z} near Prague 25068,
Czech Republic\\
$^3$ Institute of Nuclear Physics, Polish Academy of Sciences,
Cracow 31342, Poland\\
$^4$ Institut f\"ur~Theoretische Physik, Universit\"at T\"ubingen,
Kepler Center for Astro and Particle Physics,
Auf der Morgenstelle 14, D--72076 T\"ubingen, Germany\\
$^5$ Department
of Physics, Tomsk State University, 634050 Tomsk, Russia\\
$^6$ Institute of Physics, Slovak Academy of Sciences, Bratislava
84511, Slovak Republic\\}
\ead{
surovcev@theor.jinr.ru,~bydz@ujf.cas.cz,~robert.kaminski@ifj.edu.pl,
valeri.lyubovitskij@uni-tuebingen.de,
miroslav.nagy@savba.sk}

\date{\today}

\begin{abstract}

Peculiarities of obtaining parameters for broad multi-channel resonances
from data are discussed analyzing the experimental data on processes
$\pi\pi\to\pi\pi,K\overline{K}$ in the $I^GJ^{PC}=0^+0^{++}$ channel in
a model-independent approach based on analyticity and unitarity and using
an uniformization procedure. We show that it is possible to obtain a
good description of the $\pi\pi$ scattering data from the threshold to
1.89~GeV with parameters of resonances cited in the PDG tables as preferred.
However, in this case, first, representation of the $\pi\pi$ background is
unsatisfactory; second, the data on the coupled process
$\pi\pi\to K\overline{K}$
are not well described even qualitatively above 1.15~GeV when using the
resonance parameters from the only $\pi\pi$ scattering analysis.
The combined analysis of these coupled processes is needed, which is carried
out satisfactorily. Then both above-indicated flaws, related to the analysis
of solely the $\pi\pi$-scattering, are cured.
The most remarkable change of parameters with respect to the values of
only $\pi\pi$ scattering analysis appears for the mass of the $f_0 (600)$
which is now in some accordance with the Weinberg prediction on the basis
of mended symmetry and with an analysis using the large-$N_c$ consistency
conditions between the unitarization and resonance saturation.
The obtained $\pi\pi$-scattering length $a_0^0$ in case when we restrict
to the analysis of the $\pi\pi$ scattering or consider so-called
A-solution (with a lower mass and width of $f_0(600)$ meson)
agrees well with prediction of chiral perturbation theory (ChPT)
and with data extracted at CERN by the NA48/2 Collaboration from the
analysis of the $K_{e4}$ decay and  by the DIRAC Collaboration from the
measurement of the $\pi^+\pi^-$ lifetime.

\end{abstract}

\vskip 1cm



\maketitle

\section{Introduction}

The study of scalar mesons is very important for understanding the QCD vacuum.
Due the same quantum numbers of the vacuum and scalar mesons there 
are possible direct transitions between vacuum and scalars mesons. 
Therefore the study of scalar mesons can shed light on the problem 
of QCD vacuum. 
However, despite the big effort devoted to studying various aspects of
the problem~\cite{PDG} (for recent reviews see,
e.g. \cite{Amsler04,Bugg04,Close02,Klempt07}) a description of this mesonic
sector is far from being complete.
Parameters of the scalar mesons, their nature and status of some of them are
still not well settled~\cite{PDG}. For example, applying our
model-independent method in the three-channel analyses of processes
$\pi\pi\to\pi\pi,K\overline{K}$, and $\eta\eta$ or $\eta\eta^\prime$
\cite{SBKN-PRD10,SBL-arXiv11} we have obtained parameters of the $f_0(600)$
and $f_0(1500)$ which differ considerably from results of analyses which
utilize other methods (mainly those based on the dispersion relations and
Breit--Wigner approaches). 
Note, existence of the $f_0(1370)$ meson is still not obvious. 
In some works, e.g.,
in \cite{MO-02-2,Ochs10} one did not find any evidence for the existence of the
$f_0(1370)$. On the other hand, in Ref.~\cite{Bugg1370} a number of data 
requiring apparently the existence of the $f_0(1370)$ is indicated. 
We have shown \cite{SBL-arXiv11} that an existence of the $f_0(1370)$ does 
not contradict the data on processes $\pi\pi\to\pi\pi,K\overline{K},
\eta\eta(\eta\eta^\prime)$ and if this state exists,
it has a dominant $s{\bar s}$ component. In the hidden gauge unitary approach, 
the $f_0(1370)$ appears dynamically generated as a $\rho\rho$ 
state~\cite{Geng_Oset}. Also in Ref.~\cite{Geng_Oset} $f_0(1710)$ appears 
as generated from the $K^\ast \bar K^\ast$  interaction. 

This difference is very interesting and its reasons should be understood
because our method of analysis is based only on the demand for analyticity
and unitarity
of the amplitude using a uniformization procedure. The construction of the
amplitude is practically free from any dynamical (model) assumptions utilizing
only the {\it mathematical} fact that a local behavior of analytic functions
determined on the Riemann surface is governed by the nearest singularities on
all corresponding sheets. I.e., the obtained parameters of resonances can be
considered as free from theoretical prejudice. 

To better understand reasons for the difference in results, in this paper we
have performed first the two-channel model-independent analysis only of the
$\pi\pi$ scattering data. The thing is that in our previous three-channel
analysis with the uniformizing variable \cite{SBKN-PRD10,SBL-arXiv11} we were
enforced to construct a four-sheeted model of the initial eight-sheeted
Riemann surface. This we have achieved by neglecting the lowest
$\pi\pi$-threshold branch-point which means that we have considered the nearest
to the physical region semi-sheets of the initial Riemann surface. This is in
the line with our approach of a consistent account of the nearest singularities
on all relevant sheets. The two-channel analysis utilizes the full Riemann
surface and is, therefore, free of these approximations.

In the two-channel analysis only of the $\pi\pi$ scattering data we have
obtained a good description from the threshold to 1.89~GeV with values of
parameters of the $f_0$ resonances, which are well consistent with the ones
cited in the PDG tables~\cite{PDG} as preferred. However, it turned out
that the cross-section of the coupled process $\pi\pi\to K\overline{K}$ is
not well described even qualitatively above 1.15~GeV when using for the
relevant resonances the same values of parameters. The combined analysis of
these coupled processes is needed, which is carried out also satisfactorily.

In the presented combined two-channel analysis of data on processes
$\pi\pi\to\pi\pi$ and $\pi\pi\to K\overline{K}$, we have checked whether
the results of our previous three-channel
analysis \cite{SBKN-PRD10,SBL-arXiv11}
are also obtained in the two-channel case and therefore shown whether
the above-indicated assumptions are justified.
It is also interesting to determine and discuss the scattering length $a_0^0$
and slope $b_0^0$, related to the effective range of interaction, in the
separate analyses using the alternative data on $\pi\pi$ scattering from
Refs.~\cite{Hya73} and \cite{Kamin02}, adding the very precise
NA48/2-Collaboration $\pi\pi$-data~\cite{NA48/2}.
Our preliminary results have been published in Ref.~\cite{Surovtsev:2012vg}. 
In comparison with Ref.~\cite{Surovtsev:2012vg} we included additional 
combined analysis of alternative data on $\pi\pi$ scattering~\cite{Kamin02} 
and $\pi\pi \to K \bar K$. Also we included more details and explanations.   

Our paper is structured as follows.
In Sect.~II a basic formalism for our two-channel model-independent method is
shown and rules for the calculation of resonance parameters discussed. Results
of analyses are presented and discussed in Sect.~III, first only for the
$\pi\pi$ data to clarify a consistency of our results with values of parameters
from the PDG tables and then for the combined two-channel analysis, using two
sets of data for the $\pi\pi$ scattering \cite{Hya73,Kamin02} and all
accessible data~\cite{expd2} for $\pi\pi\to K\overline{K}$, to verify a
plausibility of our assumptions in the three-channel calculations.
Conclusions are provided in Sect. IV.

\section{The coupled-channel formalism in uniformizing variable method}

Our ``model-independent" method which essentially utilizes
an uniformizing variable can be used without any further assumptions
only for the two-channel case. In the three-channel case, some assumptions
about the Riemann surface have to be made~\cite{SBKN-PRD10,SBL-arXiv11}.
In this work we consider the two-channel case.

The two-channel $S$-matrix is determined on the four-sheeted Riemann
surface. The matrix elements $S_{ij}$, where
$i,j=1(\pi\pi),2(K\overline{K})$ denote channels, have the right-hand
cuts along the real axis of the $s$ complex plane ($s$ is the
invariant total energy squared), starting with the channel thresholds
$s_i$ ($i=1,2$), and the left-hand cuts related to the crossed channels.
The Riemann-surface sheets are numbered according to the signs of
analytic continuations of the roots $\sqrt{s-s_i}~~(i=1,2)$ as follows:
$\mbox{signs}\bigl(\mbox{Im}\sqrt{s-s_1},\mbox{Im}\sqrt{s-s_2}\bigl)=
++,-+,--,+-$ correspond to sheets I, II, III, IV.

The resonance representations on the Riemann surface are obtained from
the formulas which express an analytic continuation of the $S$-matrix
elements to unphysical sheets in terms of the matrix elements on sheet I
(physical) having only resonance zeros (beyond the real axis),
at least, around the physical region:
\eq\label{S_contin}
\hspace*{-1.5cm}
&&S_{11}^{\rm II}=\frac{1}{S_{11}^{\rm I}},\qquad~~~~~~~~~~~~~
S_{11}^{\rm III}=\frac{S_{22}^{\rm I}}{S_{11}^{\rm
I}S_{22}^I-(S_{12}^{\rm I})^2}, \quad
S_{11}^{\rm IV}=\frac{S_{11}^{\rm I}S_{22}^{\rm I}
-(S_{12}^{\rm I})^2}{S_{22}^I},\nonumber\\
\hspace*{-1.5cm}
&&S_{22}^{\rm II}=\frac{S_{11}^{\rm I}S_{22}^{\rm I}-(S_{12}^{\rm
I})^2}{S_{11}^{\rm I}},\quad S_{22}^{\rm III}=\frac{S_{11}^{\rm
I}}{S_{11}^{\rm I}S_{22}^{\rm I}-(S_{12}^{\rm I})^2},\quad
S_{22}^{\rm IV}=\frac{1}{S_{22}^{\rm I}},\\
\hspace*{-1.5cm}
&&S_{12}^{\rm II}=\frac{iS_{12}^{\rm I}}{S_{11}^{\rm
I}},\qquad~~~~~~~~~~~ S_{12}^{\rm III}=\frac{-S_{12}^{\rm
I}}{S_{11}^{\rm I}S_{22}^{\rm I}-(S_{12}^{\rm I})^2},\quad
S_{12}^{\rm IV}=\frac{iS_{12}^{\rm I}}{S_{22}^{\rm I}}.\nonumber
\en
Then, starting from the resonance zeros on sheet I, one can obtain
an arrangement of poles and zeros of a resonance on the whole Riemann
surface.

In the one-channel consideration of the scattering process $1\to1$
the main model-independent contribution of a resonance is given
by a pair of conjugate poles on sheet II and by a pair of conjugate
zeros on sheet I at the same conjugate points of complex energy
in $S_{11}$. In the two-channel consideration of the processes
$1\to1$ and $1\to2$, a resonance is represented by a pair of
conjugate poles on sheet II and by a pair of conjugate zeros on
sheet I in $S_{11}$ and also (as it is seen from Eq.~(\ref{S_contin}))
by a pair of conjugate poles on sheet III and by a pair of conjugate
zeros on sheet IV at the same conjugate points of complex energy
if the coupling of channels is absent ($S_{12}=0$).
If a resonance decays into both channels and/or takes part in
processes of exchange in the crossing channels, the coupling of
channels arises ($S_{12}\not=0$). Then positions of the conjugate
poles on sheet III (and of corresponding zeros on sheet IV) turn out
to be shifted with respect to the positions of the zeros on sheet I.
Thus we obtain a cluster of poles and zeros (the pole cluster of type
({\bf a})) which gives the main model-independent contribution of the
corresponding two-channel resonance. Obviously, depending on nature of
a resonance there are two more pole clusters: when the pair of conjugate
zeros on sheet I, corresponding to the resonance, is present only in
$S_{22}$ -- the pole cluster of type ({\bf b}) -- and when in each of
$S_{11}$ and $S_{22}$ --  of type ({\bf c}).
For the resonances of type ({\bf b}), the pair of complex conjugate
poles on sheet III is shifted relative to the pair of poles on sheet IV.
For the states of type ({\bf c}), one must consider the corresponding
two pairs of conjugate poles on sheet III.

It is clear that for calculating the resonance parameters (masses,
total widths and coupling constants with channels) one must use the
poles on those sheets where they are not shifted due to the coupling
of channels as they respect the zero position on the physical (I)
sheet. For resonances of types ({\bf a}) and ({\bf b}) these poles
are on sheets II and IV, respectively. For resonance of type ({\bf c})
the poles can be used on both these sheets.

Analogous consideration can be carried out in the three-channel case
\cite{KMS96,PRD-01}. Seven types of resonances arise there. Formulas
of analytic continuations of the $S$-matrix elements to unphysical
sheets in the general case of N channels can be found in \cite{KMS96}.
In these cases one can see that only on the sheets with the numbers
$2^i$ ($i=1,\cdots,N$ is the number of channel),
i.e. II, IV, VIII,$\ldots$,
the analytic continuations of the $S$-matrix elements have the form
$\propto 1/S_{ii}^{\rm I}$, where $S_{ii}^{\rm I}$ is the $S$-matrix
elements on the physical (I) sheet. This means that the pole positions
of resonances only on these sheets are at the same points of the
$s$-plane, as the resonance zeros on the physical sheet, and are not
shifted due to the coupling of channels. Therefore, {\it the resonance
parameters should be calculated from the pole positions only on these
sheets}. It seems that neglecting this fact serious misunderstandings
can arise.  This concerns analyses which do not consider the structure
of the Riemann surface of the $S$-matrix and especially the analyses
of the decay processes in which, as a rule, the multi-channel nature
of resonances is not taken into account.

In the combined analysis of coupled processes, it is convenient to use
the Le Couteur--Newton relations~\cite{LeCou}. They express the
$S$-matrix elements of all coupled processes in terms of the Jost
matrix determinant $d(\sqrt{s-s_1},\cdots,\sqrt{s-s_N})$ that is a
real analytic function with the only square-root branch-points at
$\sqrt{s-s_\alpha}=0$.

A necessary and sufficient condition for existence of the multi-channel
resonance is its representation by one of the types of pole clusters. In
order to use the representation of resonances by the pole clusters which
is very important for the wide multi-channel states, we take advantage
of the fact that the amplitude is a one-valued function on the Riemann
surface. To this end, a uniformizing variable is applied, which maps
the Riemann surface onto a plane.

\begin{figure}[htb]
\begin{center}
\includegraphics[width=0.38\textwidth,angle=0]{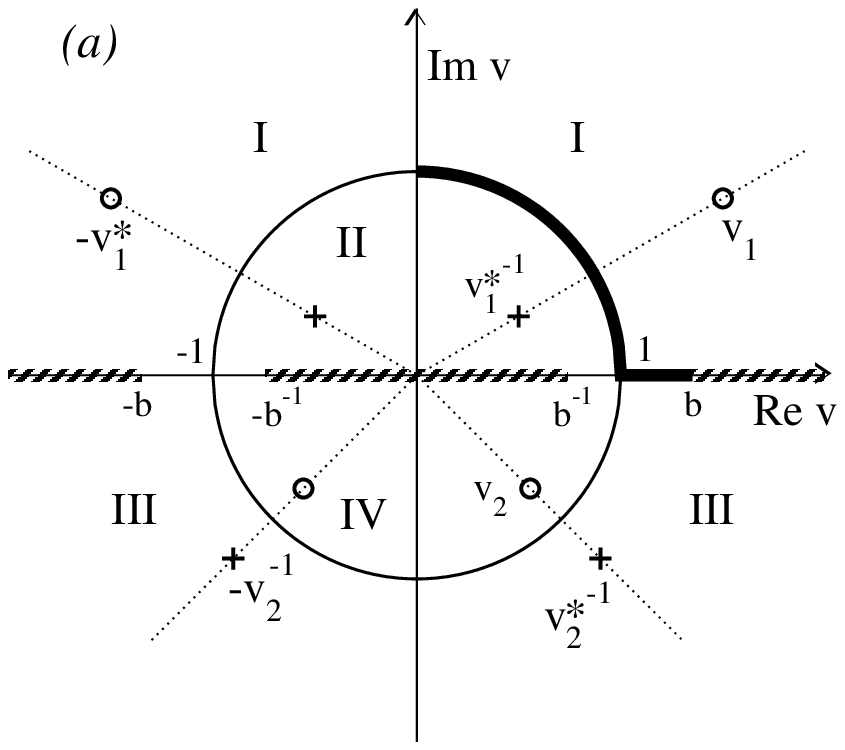}
\includegraphics[width=0.38\textwidth,angle=0]{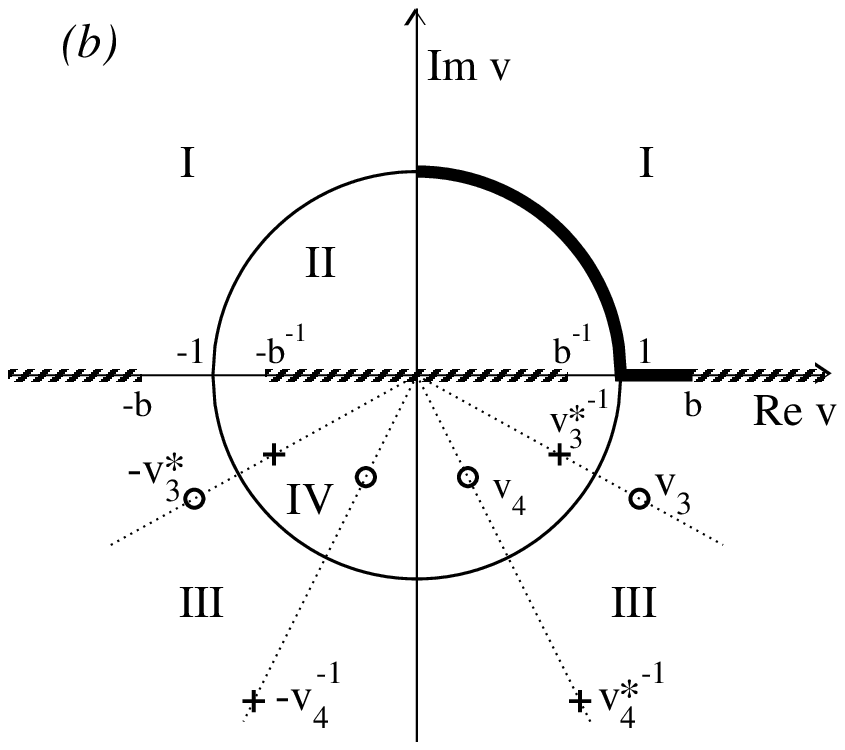}
\includegraphics[width=0.38\textwidth,angle=0]{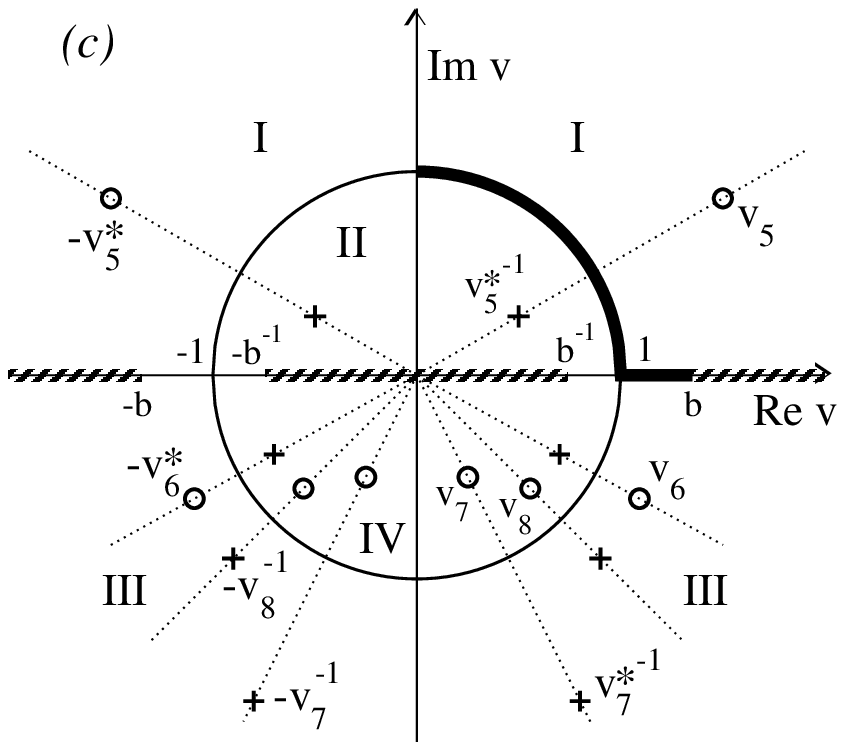}
\caption{Representation of resonances of type ({\bf a}),
({\bf b}) and ({\bf c}) on the uniformization $v$-plane in
$S_{11}$.\label{fig:lzplane}}
\end{center}
\end{figure}

In this two-channel analysis of processes $\pi\pi\to\pi\pi,K\overline{K}$
we applied the uniformizing variable which takes into account, in
addition to the $\pi\pi$- and $K\overline{K}$-threshold branch-points,
the left-hand branch-point at $s=0$, related to the $\pi\pi$ crossed
channels:
\eq\label{lv}
v=\frac{m_K\sqrt{s-4m_\pi^2}\,+\,
m_\pi\sqrt{s-4m_K^2}}{\sqrt{s(m_K^2-m_\pi^2)}}\,.
\en
It maps the four-sheeted Riemann surface with two unitary cuts and
the left-hand cut onto the $v$-plane \cite{SKN-epja02} divided into
two parts by a unit circle centered at the origin.
In Figure~\ref{fig:lzplane}, an uniformization $v$-plane for the
two-channel-$\pi\pi$-scattering amplitude is shown with the representation
of resonances of types ({\bf a}), ({\bf b}) and ({\bf c}):
the Roman numerals (I,\ldots, IV) denote the images of the
corresponding sheets; the thick line represents the physical region; the
points $i$, 1 and $b=\sqrt{(m_K+m_\pi)/(m_K-m_\pi)}$ correspond to the
$\pi\pi, K\overline{K}$ thresholds and $s=\infty$, respectively;
the shaded intervals ~$(-\infty,-b],~[-b^{-1},~b^{-1}],~[b,\infty)$~ are
the images of the corresponding edges of the left-hand cut.
The depicted positions of poles ($+$) and of zeros ($\circ$) give
the resonance representations of the type ({\bf a}), ({\bf b}) and ({\bf c})
in $S_{11}$.
%
%
The resonance poles are symmetric to the corresponding zeros with respect
to the unit circle that guarantees the elastic unitarity of $\pi\pi$
scattering up to the $K\overline{K}$ threshold. The whole picture of
poles and zeros is symmetric with respect to the imaginary axis
guaranteeing a property of the real analyticity of the $S$-matrix.

On the $v$-plane, $S_{11}(v)$ has no cuts; $S_{12}(v)$ and $S_{22}(v)$ do
have the cuts which arise from the left-hand cut on the $s$-plane,
starting at $s=4(m_K^2-m_\pi^2)$, which is further approximated by a pole
\eq
d_L&=&v^{-4} \biggl(1-(p-i\sqrt{1-p^2})v)^4(1+(p+i\sqrt{1-p^2})v\biggr)^4
\nonumber\\
& &({\rm from~analysis}~~ p=0.903\pm0.0004)\,.
\label{eq:d_L}
\en
The fourth power is stipulated by the following model-independent
arguments. First, a pole on the real $s$-axis on the physical sheet in
$S_{22}$ is accompanied by a pole on sheet II at the same $s$-value
(see Eq.~(\ref{S_contin})). On the $v$-plane this implies the pole of
second order (and also zero of the same order, symmetric to the pole
with respect to the real axis). Second, for the $s$-channel process
$\pi\pi\to K\overline{K}$, the crossing $u$- and $t$-channels are the
$\pi-K$ and $\overline{\pi}-K$ scattering (exchanges in these channels
give contributions on the left-hand cut). This results in the additional
doubling of the multiplicity of the indicated pole on the $v$-plane.
Therefore, the contribution of the left-hand branch-point at
$s=4(m_K^2-m_\pi^2)$ is approximated as the fourth-power pole on the
real $s$-axis on the physical sheet in the sub-$K\overline{K}$-threshold
region.

On the $v$-plane, the Le Couteur-Newton relations are \cite{LeCou,SKN-epja02}
\eq
S_{11}=\frac{d(-v^{-1})}{d(v)},\quad S_{22}=\frac{d(v^{-1})}{d(v)},
\quad S_{11}S_{22}-S_{12}^2=\frac{d(-v)}{d(v)}.
\en
The main model-independent contribution of resonances, given by the pole
clusters, is factorized in the $S$-matrix elements from the background.
The possible remaining small (model-dependent) contributions of
resonances are supposed to be included in the background. Therefore
the $d(v)$-function, which does not possess already branch points, is taken
as
\eq
d=d_{res}d_Ld_{bg}\,.
\en
The function $d_{res}(v)$ represents the contribution of resonances,
described by one of three types of the pole-zero clusters, {\it i.e.},
\eq
d_{res} = v^{-M}\prod_{n=1}^{M} (1-v_n^* v)(1+v_n v)\,,
\en
where $M$ is the number of pairs of the conjugate zeros.

The background part is
\eq
d_{bg}=\mbox{exp}\biggl[
-i\sum_{n=1}^{3}\frac{\sqrt{s-s_n}}{2m_n}(\alpha_n+i\beta_n)
\biggr]
\en
with
\eq
\hspace*{-1.5cm}
\alpha_n &=& a_{n1}+a_{n\eta}\frac{s-s_\eta}{s_\eta}\theta(s-s_\eta)
+a_{n\sigma}\frac{s-s_\sigma}{s_\sigma}\theta(s-s_\sigma)+
a_{nv}\frac{s-s_v}{s_v}\theta(s-s_v)\,, \nonumber\\
\hspace*{-1.5cm}
\beta_n&=& b_{n1}+b_{n\eta}\frac{s-s_\eta}{s_\eta}\theta(s-s_\eta)
+b_{n\sigma}\frac{s-s_\sigma}{s_\sigma}\theta(s-s_\sigma)+
b_{nv}\frac{s-s_v}{s_v}\theta(s-s_v)\,,
\en
where $s_\eta$ and $s_\sigma$ are the $\eta\eta$ and $\sigma\sigma$
thresholds, respectively (the latter should be determined in the
analysis), $s_v$ is a combined threshold of many opened channels in
the vicinity of 1.5~GeV (e.g.,
$\eta\eta^{\prime},~\rho\rho,~\omega\omega$) and it is determined in
the analysis: $s_\sigma=1.6558~{\rm GeV}^2$, $s_v=2.1293~{\rm
GeV}^2$.

The data used in the analysis are the results of phase analyses which
are given for phase shifts of the amplitudes $\delta_{\alpha\beta}$
and for modules of the $S$-matrix elements
$\eta_{\alpha\beta}=|S_{\alpha\beta}|$ ($\alpha,\beta=1,~2$)
\eq
S_{\alpha\alpha}=\eta_{\alpha\alpha}e^{2i\delta_{\alpha\alpha}}\,,~~~~~
S_{\alpha\beta}=\eta_{\alpha\beta}e^{i\phi_{\alpha\beta}}\,.
\en
The two-channel unitarity provides the relations
\eq
\eta_{11}=\eta_{22}\,, ~~
\eta_{12}=\sqrt{1-{\eta_{11}}^2}\,,~~
\phi_{12}=\delta_{11}+\delta_{22}\,.
\label{eq:2-ch-uni}
\en

\section{Analysis of the data on isoscalar S-wave processes
$\pi\pi\to\pi\pi,K\overline{K}$}

We analyzed data on processes ~$\pi\pi\to\pi\pi,K\overline{K}$. For the
$\pi\pi$ scattering, we took alternative data -- {\bf set~I:} for
$0.575~{\rm GeV} < \sqrt{s} < 1.89~{\rm GeV}$ from Ref.~\cite{Hya73}
and for $\sqrt{s} < 1~{\rm GeV}$ from Refs.~\cite{NA48/2,expd1,%
expd5,expd6,Rosselet77,Bel'kov79};
{\bf set~II:} for $0.61~{\rm GeV} < \sqrt{s} < 1.59~{\rm GeV}$ from
Refs.~\cite{Kamin02} and \cite{Grayer74} (solution G) and for
$\sqrt{s} < 1~{\rm GeV}$ as in set~I. For $\pi\pi\to K\overline{K}$, we
used practically all accessible data~\cite{expd2}.

Initially analyzing only the $\pi\pi$ scattering from set~I, we
demonstrated that it is possible to achieve an excellent description of
the data for the phase shift $\delta_{11}$ and modulus of the $S$-matrix
element (the total $\chi^2/\mbox{NDF}=168.433/(189-33)\approx1.08$) with
the parameters of resonances (Table \ref{tab:pole_clusters_pipi})
which largely coincide with the values cited as estimation of the PDG
\cite{PDG}, though a negative phase-shift in the background on the
$\pi\pi$ threshold arises.
%
%
\begin{table}[htb!]
\caption{Pole clusters for resonances on the complex energy plane
$\sqrt{s}$ in the analysis of only $\pi\pi$-scattering without and
with the narrow $f_0(1500)$. Pole energies
~$\sqrt{s_r}\!=\!{\rm E}_r\!-\!i\Gamma_r/2$ in MeV are shown.}
\label{tab:pole_clusters_pipi}
\begin{center}
{
\vskip-0.2truecm
\def\arraystretch{1.2}
\begin{tabular}{|c|c|c|c|c|}
\hline \multicolumn{5}{|c|}{Without the narrow $f_0(1500)$} \\
\hline \multicolumn{2}{|c|}{Sheet} & II & III & IV \\ \hline
{$f_0(600)$} & {${\rm E}_r$} &  447.5$\pm$5.4 & 492.6$\pm$31.5 & {} \\
{}           & {$\Gamma_r/2$}&  267.0$\pm$6.2 & 307.8$\pm$15.6 & {} \\ \hline
{$f_0(980)$} & {${\rm E}_r$} & 1001.1$\pm$3.4 & 978.5$\pm$ 9.5 & {} \\
{}           & {$\Gamma_r/2$}&   20.3$\pm$2.4 &  38.5$\pm$ 6.3 & {} \\ \hline
{$f_0(1370)$}& {${\rm E}_r$} & {}  & 1382.6$\pm$38.5 & 1301.2$\pm$38.2 \\
{}           & {$\Gamma_r/2$}& {}  &  179.5$\pm$39.2
&  243.0$\pm$52.8 \\ \hline
{$f_0(1500)$}& {${\rm E}_r$} & 1512.0$\pm$10.6
& 1499.0$\pm$108.3~~~1509.1$\pm$112.6
             & 1505.9$\pm$30.0 \\
{}           & {$\Gamma_r/2$} & 191.0$\pm$10.5
&  310.2$\pm$ 71.8~~~241.0$\pm$68.9
             & 168.0$\pm$32.8 \\ \hline
{$f_0(1710)$}& {${\rm E}_r$} & {} & 1700.3$\pm$30.5 & 1720.1$\pm$30.5 \\
{}           & {$\Gamma_r/2$}& {} &   58.8$\pm$29.5 &   64.9$\pm$34.3 \\
\hline \multicolumn{5}{|c|}{With the narrow $f_0(1500)$} \\
\hline \multicolumn{2}{|c|}{Sheet} & II & III & IV \\ \hline
{$f_0(600)$} & {${\rm E}_r$} & 447.5$\pm$5.9 & 492.7$\pm$36.0 & {} \\
{}           & {$\Gamma_r/2$}& 267.0$\pm$6.5 & 307.8$\pm$16.5 & {} \\ \hline
{$f_0(980)$} & {${\rm E}_r$} & 1001.1$\pm$3.7 & 979.1$\pm$12.0 & {} \\
{}           & {$\Gamma_r/2$}& 20.3$\pm$2.6 & 38.5$\pm$7.1 & {} \\ \hline
{$f_0(1370)$}& {${\rm E}_r$} & {} & 1375.8$\pm$51.5 & 1301.1$\pm$47.9 \\
{}           & {$\Gamma_r/2$}& {} &  179.5$\pm$36.5 &  224.0$\pm$49.3 \\ \hline
{$f_0(1500)$}& {${\rm E}_r$} & {} & 1498.8$\pm$39.3 & 1503.7$\pm$45.1 \\
{}           & {$\Gamma_r/2$}& {} &   51.8$\pm$43.3 &   56.5$\pm$39.4 \\ \hline
{$f_0^\prime(1500)$} & {${\rm E}_r$} & 1511.4$\pm$11.2 & 1499.8$\pm$104.3~~~
               1509.1$\pm$119.4 & 1505.9$\pm$38.5 \\
{}           & {$\Gamma_r/2$} & 200.5$\pm$11.0
&  310.5$\pm$58.8~~~241.0$\pm$63.8
             &  168.1$\pm$40.6 \\ \hline
{$f_0(1710)$}& {${\rm E}_r$} & {} & 1700.3$\pm$31.2 & 1720.1$\pm$32.2 \\
{}           & {$\Gamma_r/2$}& {} &   58.6$\pm$26.4 &   64.9$\pm$30.1 \\ \hline
\end{tabular}}
\end{center}
\end{table}
E.g., for the $f_0(600)$  the found pole on
sheet II coincides practically with the one at around $450 - i275$~MeV
which was found in the recent dispersive $\pi\pi$-scattering data
analyses~\cite{Colangelo01,Garcia11}.
We would like to stress an agreement
of our prediction for the mass of $f_0(980)$ meson 1001~MeV
with PDG result 990$\pm$20~MeV.
Note, the mass of $f_0(980)$ slightly above 1~GeV was also obtained in many
other works which analyzed the $\pi\pi$ scattering
(e.g.~\cite{GarciaMKPRE-11}). Only a difference with PDG occurs
for the width of $f_0(1500)$ (336~MeV against 109$\pm$7~MeV of PDG).
We think that the observed wide resonance $f_0(1500)$
is in reality a superposition of two states, wide and narrow. The narrow
state is observed in processes considered in works cited by the PDG.
To test this interpretation of the $f_0(1500)$, we analyzed the
$\pi\pi$-scattering data also assuming both wide and narrow $f_0(1500)$.
The gained description is excellent:
the total $\chi^2/\mbox{NDF}=171.715/(189-29)\approx1.07$.
The obtained parameters of resonances are shown in
Table~\ref{tab:pole_clusters_pipi}. Now the parameters of the narrow
$f_0(1500)$ are consistent with those in the PDG tables. In the
presented analyses, the $f_0 (600)$ and $f_0 (980)$ are described by
the clusters of type ({\bf a}); $f_0 (1370)$, $f_0(1500)$ and
$f_0(1710)$, type ({\bf b}); $f_0^\prime(1500)$, type ({\bf c}). The
received background parameters in the analysis with the narrow
$f_0(1500)$ are:
$a_{11}=-0.0895\pm0.0030$,  $a_{1\eta}=0.04\pm0.03$,
$a_{1\sigma}=0.0\pm0.8$,    $a_{1v}=0.0\pm0.7$,
$b_{11}=0.0\pm0.007$, $b_{1\eta}=0.0\pm0.01$,
$b_{1\sigma}=0.0\pm0.02$, $b_{1v}=0.054\pm0.036$. 
The number of parameters is determined by that each resonance 
is described by pole clusters (it can be seen from the formulas  
for analytical continuations). There are also parameters related 
with opened channels in the background. 

\begin{figure}[htb]
\begin{center}
\includegraphics[width=0.49\textwidth,angle=0]{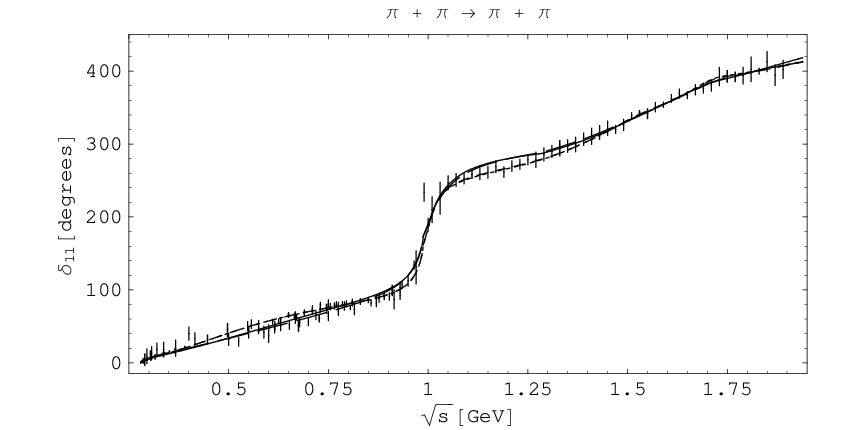}
\includegraphics[width=0.49\textwidth,angle=0]{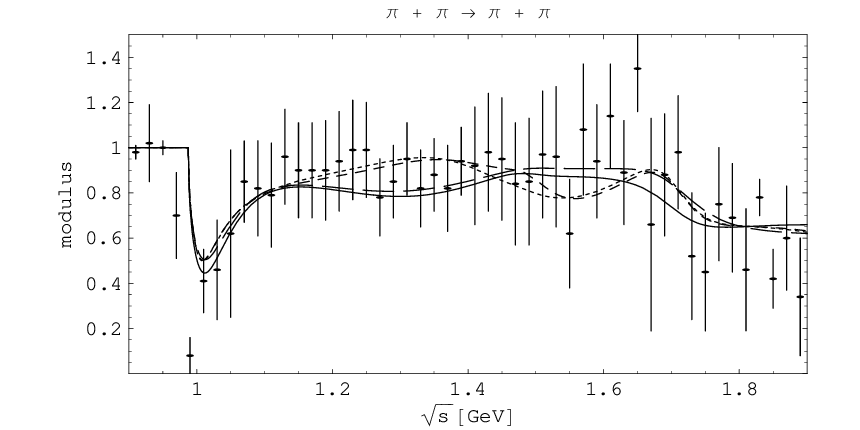}\\
\includegraphics[width=0.49\textwidth,angle=0]{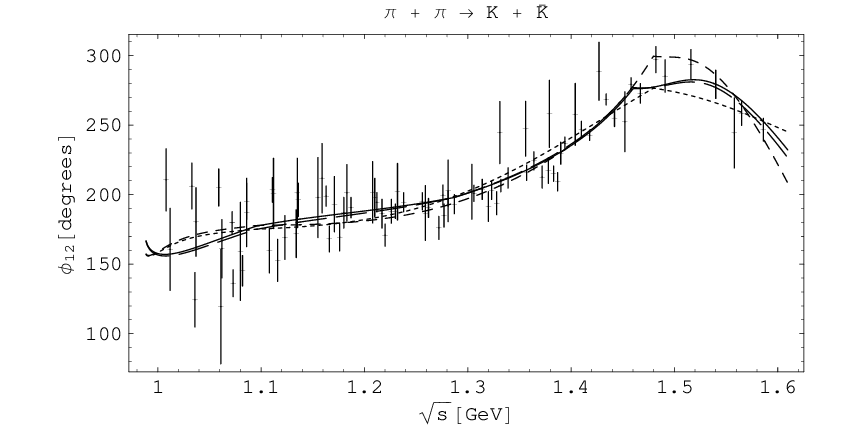}
\includegraphics[width=0.49\textwidth,angle=0]{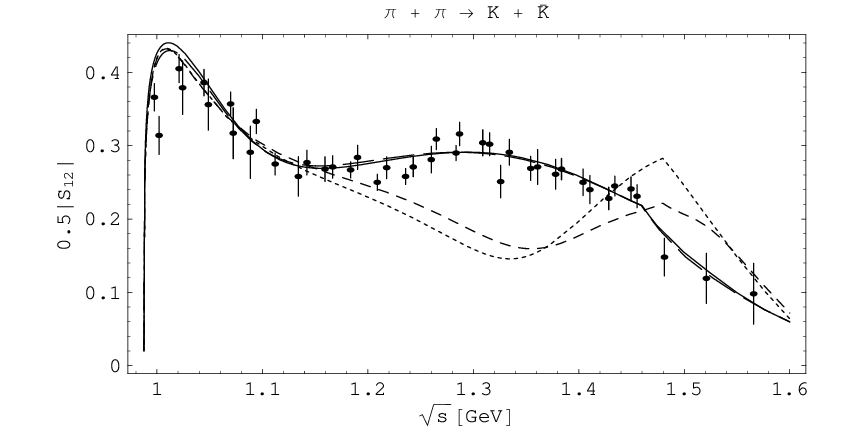}
\caption{The $S$-wave phase shifts and modules of the $\pi\pi$-scattering
matrix element (upper row) and of the $\pi\pi\to K\overline{K}$ matrix
element in analyses of set I.
The dotted and short-dashed lines correspond to the analysis of only
$\pi\pi$-scattering without and with the narrow $f_0(1500)$, respectively.
The long-dashed and solid lines correspond to solutions A and B of the
combined analysis of $\pi\pi\to\pi\pi,K\overline{K}$, respectively.
The data are from
Refs.~\cite{Hya73,NA48/2,expd2,expd1,expd5,expd6,Rosselet77,Bel'kov79}.}
\label{fig:set_I}
\end{center}
\end{figure}

In Figure~\ref{fig:set_I} results of the fitting only to the
$\pi\pi$-scattering data are shown (upper row); in the lower row
there are given energy behaviors of the phase shift and module of
the $\pi\pi\to K\overline{K}$ matrix element which are calculated
using the resonance parameters from the analysis of only
$\pi\pi$-scattering: the dotted and short-dashed lines correspond
to the analysis without and with the narrow $f_0(1500)$, respectively.
%
%
The gained description of data is very good. The $\pi\pi$ scattering
length $a_0^0$, obtained in these analyses of the data from set I, is
$0.222\pm 0.008~m_{\pi^+}^{-1}$, which is also in the very good agreement
with the experimental results and with the results of the ChPT
calculations (see Tab.~V). However, let us also emphasize two important
flaws:
\begin{itemize}
\item
First, the negative phase-shift in the background beginning from the
$\pi\pi$ threshold ($a_{11}=-0.0895$) seems to be necessary for a
successful description of the data. This should not be the case because,
in the uniformizing variable, we have taken into account the left-hand
branch-point at $s=0$ which gives a main contribution to the $\pi\pi$
background below the $K\overline{K}$ threshold. Other possible
contributions of the left-hand cut related with exchanges by the nearest
mesons -- the $\rho$-meson and the $f_0 (600)$ -- practically obliterate
each other \cite{SKN-epja02} because vector and scalar particles
contribute with the opposite signs due to the gauge invariance.
\item
Second, the description of the data on reaction
$\pi\pi\to K\overline{K}$, using the same parameters of resonances as
in the $\pi\pi$ channel, is satisfactory only for the phase shift
$\phi_{12}$ which is due to the approximation of the left-hand
branch-point at $s=4(m_K^2-m_\pi^2)$ in $S_{12}$ and $S_{22}$ by the
fourth-power pole. The module of the $S$-matrix element $\eta_{12}$ is
described well only from the $K\overline{K}$ threshold up to about 1.15~GeV
as it should be due to the two-channel 
unitarity (see Eqs.~(\ref{eq:2-ch-uni})).
Above this energy the description fails even qualitatively
(see Fig.~\ref{fig:set_I}).
\end{itemize}
From this we conclude that: If the data are consistent, for obtaining
correct parameters of wide resonances the combined analysis of data
on coupled processes is needed. Further the combined analyses of data
(sets I and II) on processes $\pi\pi\to\pi\pi,K\overline{K}$ are performed
supposing that in the 1500-MeV region there are two resonances. Here it
ought to be noted that the consideration of both sets of data on
the $\pi\pi$ scattering is needed in the combined analyses because these
data differ from each other in energy ranges 1.55-1.85 and 1.25-1.55~GeV
especially important for this investigation.

In the analysis of set I, the resonances are described by pole clusters of
the same types as in the analysis only of $\pi\pi$-scattering. Satisfactory
combined description of two analyzed processes is obtained with the total
$\chi^2/\mbox{NDF}=391.299/(312-40)\approx1.44$. One sees that the data for
the $\pi\pi$ scattering below 1~GeV admit two solutions for the phase shift:
A and B which mutually differ mainly in the pole position on sheet II for
the $f_0(600)$. The $\chi^2$ shown above is for the solution B.
The A-solution gives a slightly worse result:
the total $\chi^2/\mbox{NDF}=416.887/(312-40)\approx1.53$.

In the analysis of set II, the resonances are described by the pole clusters
of the same types as in previous case except for the narrow $f_0(1500)$ which
is represented now by the cluster of type ({\bf a}) to provide more rapid 
growth of the $\pi\pi$ phase shift above 1.45~GeV than in the first case.

Also in this case, a satisfactory description is obtained with the total
$\chi^2/\mbox{NDF}=418.268/(306-41)\approx1.58$ for the A-solution and
$\chi^2/\mbox{NDF}=375.160/(306-41)\approx1.42$ for the B-solution.

\begin{figure}[htb]
\begin{center}
\includegraphics[width=0.49\textwidth,angle=0]{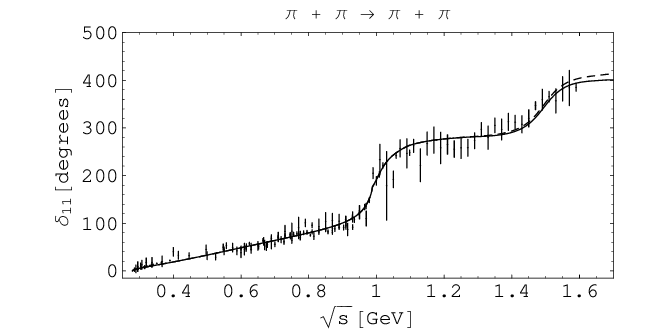}
\includegraphics[width=0.49\textwidth,angle=0]{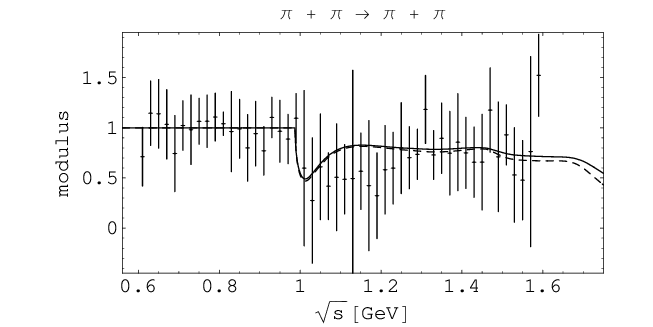}\\
\includegraphics[width=0.49\textwidth,angle=0]{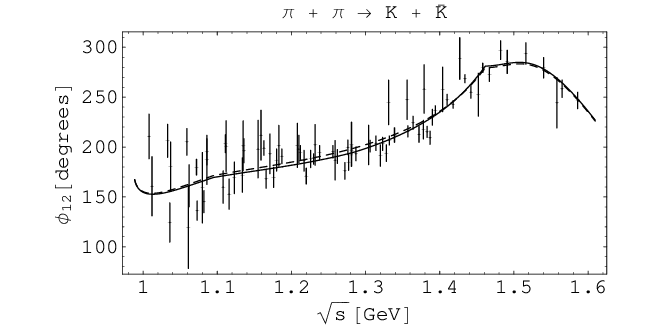}
\includegraphics[width=0.49\textwidth,angle=0]{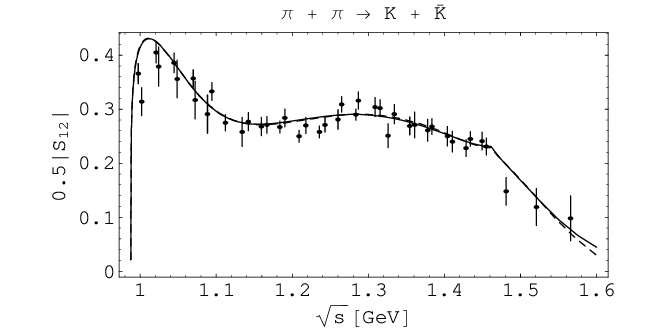}
\caption{The $S$-wave phase shifts and modules of the
$\pi\pi$-scattering matrix element (upper row) and of the
$\pi\pi\to K\overline{K}$ matrix element in the combined analyses
of the data on these two processes from set II. The solid lines
correspond to the A-solution and the dashed ones to the B-solution.
The data are from
Refs.~\cite{Kamin02,NA48/2,expd2,expd1,expd5,expd6,Grayer74,
Rosselet77,Bel'kov79}.}\label{fig:set_II}
\end{center}
\end{figure}

In Figure~\ref{fig:set_II} results of the fitting to the experimental data
from sets~II are shown. The solid lines correspond to the A-solution and
the dashed ones to the B-solution.

We note that in comparison with results of 
the dispersion-relation approach~\cite{Colangelo01,Garcia11} 
we achieved better description of both sets (I and II) of 
the $\pi\pi$-scattering data in the considerably larger energy region. 
This is a consequence of the fact that in our case the $\pi\pi$-scattering 
amplitude is taken on the 4-sheeted Riemann surface due to an explicit 
allowance for the $K\overline{K}$ threshold in the uniformizing variable, 
while in the dispersion-relation approach the 
amplitude is considered on the 2-sheeted Riemann surface. 
A price of this is that we can not take exactly into account the crossing 
symmetry of the $\pi\pi$ scattering. Contribution of the crossed processes 
of the $\pi\pi$ scattering is allowed for in part (but considerably) 
in view of taking into account the left-hand branch-point at $s=0$. 
On the other hand, we acquire that important and sensitive criterion 
of plausibility of the description of data as a reasonable and simple
representation of the background. This criterion has pushed us to carry 
out the combined analyses of data (sets I and II) on processes 
$\pi\pi\to\pi\pi,K\overline{K}$.

In Table \ref{tab:pole_clusters_A_I} we show the pole clusters for resonances
on the complex energy plane $\sqrt{s}$ for the A- and B-solutions in the
combined analyses of data (set~I) on processes $\pi\pi\to\pi\pi,K\overline{K}$.
\begin{table}[htb!]
\caption{Pole clusters for resonances on the $\sqrt{s}$-plane
in the combined analyses of data (set~I) on processes
$\pi\pi\to\pi\pi,K\overline{K}$.
~~$\sqrt{s_r}\!=\!{\rm E}_r\!-\!i\Gamma_r/2$ in MeV are shown.}
\label{tab:pole_clusters_A_I}
\begin{center}
{
\vskip-0.2truecm
\def\arraystretch{1.2}
\begin{tabular}{|c|c|c|c|c|}
\hline \multicolumn{5}{|c|}{A-solution}\\
\hline \multicolumn{2}{|c|}{Sheet} & II & III & IV \\ \hline
{$f_0(600)$} & {${\rm E}_r$} &
517.0$\pm$7.8 & 458.5$\pm$14.7 & {} \\
{} & {$\Gamma_r/2$} & 393.9$\pm$6.0 & 205.9$\pm$4.7 & {} \\
\hline {$f_0(980)$} & {${\rm E}_r$} &
1004.6$\pm$3.9 & 995.5$\pm$10.1 & {} \\
{} & {$\Gamma_r/2$} & 25.0$\pm$2.3 & 96.9$\pm$2.7 & {} \\ \hline
{$f_0(1370)$} & {${\rm E}_r$}
& {} & 1351.5$\pm$32.5 & 1342.9$\pm$12.2 \\
{} & {$\Gamma_r/2$} & {} & {369.0$\pm$45.7} & {221.6$\pm$30.7} \\ \hline
{$f_0(1500)$} & {${\rm E}_r$} & {} & 1498.7$\pm$5.8 & 1501.1$\pm$6.4 \\
{} & {$\Gamma_r/2$} & {} & 56.7$\pm$5.6 & 56.6$\pm$6.0 \\ \hline
{$f_0^\prime(1500)$} & {${\rm E}_r$} &
1532.2$\pm$12.4 & 1489.1$\pm$16.2~~~1515.9$\pm$29.2 & 1519.3$\pm$18.7 \\
{} & {$\Gamma_r/2$} & 323.2$\pm$21.0 & 217.9$\pm$10.2~~~388.4$\pm$22.6
& 339.5$\pm$42.2  \\ \hline
{$f_0 (1710)$} & {${\rm E}_r$} &
{} & {1701.9$\pm$31.8} & {1717.0$\pm$34.9} \\
{} & {$\Gamma_r/2$} & {} & {77.8$\pm$18.0} & {72.9$\pm$16.2} \\
\hline \multicolumn{5}{|c|}{B-solution}\\
\hline \multicolumn{2}{|c|}{Sheet} & II & III & IV \\ \hline
{$f_0(600)$} & {${\rm E}_r$} &
550.6$\pm$9.0 & 664.5$\pm$12.1 & {} \\
{} & {$\Gamma_r/2$} & 502.1$\pm$7.2 & 188.2$\pm$2.6 & {} \\
\hline {$f_0(980)$} & {${\rm E}_r$} &
1003.2$\pm$3.0 & 995.4$\pm$7.3 & {} \\
{} & {$\Gamma_r/2$} & 28.9$\pm$2.0 & 96.7$\pm$2.7  & {} \\ \hline
{$f_0(1370)$} & {${\rm E}_r$} & {} &
1353.8$\pm$27.9 & 1336.7$\pm$14.1 \\
{} & {$\Gamma_r/2$} & {} & {367.4$\pm$37.4} & {251.9$\pm$27.5} \\
\hline {$f_0(1500)$} & {${\rm E}_r$} & {} &
1499.5$\pm$6.0 & 1500.3$\pm$6.3 \\
{} & {$\Gamma_r/2$} & {} & 56.5$\pm$6.1 & 57.0$\pm$6.4 \\
\hline {$f_0^\prime(1500)$} & {${\rm E}_r$} &
1528.4$\pm$12.5 & 1491.3$\pm$15.8~~~1510.8$\pm$29.1 & 1515.6$\pm$17.0 \\
{} & {$\Gamma_r/2$} & 328.0$\pm$20.2
& 217.9$\pm$8.0~~~388.3$\pm$16.3 & 340.3$\pm$34.9\\
\hline {$f_0 (1710)$} & {${\rm E}_r$} &
{} & {1703.1$\pm$31.5} & {1722.0$\pm$35.7} \\
{} & {$\Gamma_r/2$} & {} & {81.7$\pm$19.9} & {92.3$\pm$20.3} \\
\hline
\end{tabular}}
\end{center}
\end{table}
The obtained background parameters in the analysis of set~I for the
A-solution are: $a_{11}=0.0\pm0.003$, $a_{1\eta}=-0.1004\pm0.0301$,
$a_{1\sigma}=0.2148\pm0.0822$, $a_{1v}=0.0\pm0.07$,
$b_{11}=b_{1\eta}=b_{1\sigma}=0$, $b_{1v}=0.012\pm0.0287$,
$a_{21}=-0.919\pm0.107$, $a_{2\eta}=-1.399\pm0.348$,
$a_{2\sigma}=0.0\pm0.7$, $a_{2v}=-11.45\pm0.75$,
$b_{21}=0.0747\pm0.0503$, $b_{2\eta}=b_{2\sigma}=0$,
$b_{2v}=4.83\pm1.94$;\\
for B-solution: $a_{11}=0.0\pm0.003$, $a_{1\eta}=-0.0913\pm0.0327$,
$a_{1\sigma}=0.1707\pm0.0899$, $a_{1v}=0.0\pm0.07$,
$b_{11}=b_{1\eta}=b_{1\sigma}=0$, $b_{1v}=0.006\pm0.029$,
$a_{21}=-1.338\pm0.111$, $a_{2\eta}=-1.119\pm0.376$,
$a_{2\sigma}=0.0\pm0.8$, $a_{2v}=-12.13\pm0.77$,
$b_{21}=0.018\pm0.050$, $b_{2\eta}=b_{2\sigma}=0$,
$b_{2v}=4.48\pm1.98$.

It is apparent that in the combined analysis of data on coupled
processes both above-indicated important flaws, which related
to the analysis of only $\pi\pi$-scattering, are cured.
Now the $\pi\pi$ background below the $K\overline{K}$ threshold
is absent ($a_{11}=0.0$) because its contribution is practically
completely accounted for by the left-hand branch-point at $s=0$
which is included explicitly in the uniformizing variable (\ref{lv}).
An arising pseudo-background at the $\eta\eta$-threshold
($a_{1\eta}<0$) is also clear: this is a direct indication
to consider explicitly the $\eta\eta$-threshold branch-point.
This was already done in our previous work \cite{SBL-arXiv11}.

Considering Tables~\ref{tab:pole_clusters_pipi} 
and~\ref{tab:pole_clusters_A_I},
one can see that the poles on sheet II for the $f_0(600)$ are located 
always nearer to the real $s$-axis in the first case than in the second.
This can be explained: If an analytical function,
having two important branch-points, is considered on the 2-sheeted Riemann
surface (i.e. neglecting the 2nd branch-point), the 3rd sheet can be
thought as amalgamated with the 2nd sheet (and the 1rst one with the 4th).
Therefore, if the initial function is described by a pole on sheet II and
the same pole on sheet III, then the approximated function, considered on
the 2-sheeted Riemann surface, should be described by two poles on the
2nd sheet because the pole from sheet III is turned out to be on sheet II.
Obviously, if we consider instead this last function some approximated
function, described the one pole on sheet II, then this pole will be
settled down always nearer to the real axis than the pole on the 2nd sheet
when this function was considered on the 4-sheeted Riemann surface.
I.e., in the dispersion-relation approach, one obtained some effective
pole representing a resonance. 

Generally, wide multi-channel states are most adequately represented by
pole clusters, {\it i.e.}, by the poles on all the corresponding sheets,
because the pole clusters give a main effect of resonances.
The pole positions are rather stable characteristics for various models,
whereas masses and widths are very model-dependent for wide resonances
(see a discussion in Ref. \cite{SKN-epja02}).
Earlier one noted that the wide resonance parameters are largely controlled
by the nonresonant background (see, {\it e.g.} \cite{Achasov-Shest}). In part
this problem is removed due to allowing for the left-hand branch-point at $s=0$
in the uniformizing variable. There remains only a considerable dependence of
resonance masses and widths on the used model.
E.g., if for the resonance part of the amplitude one use the form
\begin{equation}
T^{res}=\frac{\sqrt{s}\Gamma_{el}}{m_{res}^2-s-i\sqrt{s}\Gamma_{tot}}
\end{equation}
then masses and total widths (Tables \ref{tab:mass_width_I}
and \ref{tab:mass_width_II}) can be calculated using formulas
\begin{equation} \label{mass_width}
m_{res}=\sqrt{{\rm E}_r^2+\left(\frac{\Gamma_r}{2}\right)^2}~~~
{\rm and}~~~\Gamma_{tot}=\Gamma_r.
\end{equation}
where pole parameters $\!\sqrt{s_r}={\rm E}_r-i\Gamma_r/2\!$ must correspond
to the pole positions on sheets II and IV for the resonances of type ({\bf a}),
({\bf c}) and ({\bf b}), respectively.

\begin{table}[htb]
\caption{Masses and total widths of the resonances, obtained in the
analysis of set I.} \label{tab:mass_width_I}
\begin{center}
{
\vskip-0.2truecm
\def\arraystretch{1.2}
\begin{tabular}{|c|c|c|c|c|} \hline {} &
\multicolumn{2}{|c|}{A-solution}& \multicolumn{2}{|c|}{B-solution} \\\hline
{State} & $m_{res}$[MeV] & $\Gamma_{tot}$[MeV] & $m_{res}$[MeV]
& $\Gamma_{tot}$[MeV]
\\ \hline $f_0 (600)$
&  650.0$\pm$9.8  & 787.8$\pm$12.0  &  745.2$\pm$11.5 & 1004.2$\pm$14.4 \\
\hline $f_0 (980)$
& 1004.9$\pm$4.0  &  50.0$\pm$4.6   & 1003.6$\pm$3.1  & 57.8$\pm$4.0 \\
\hline $f_0 (1370)$
& 1361.1$\pm$17.0 & 443.2$\pm$61.4  & 1360.2$\pm$18.9 & 503.8 $\pm$55.0 \\
\hline $f_0(1500)$
& 1502.2$\pm$6.6  & 113.2$\pm$12.0  & 1501.4$\pm$6.5  & 114.0$\pm$12.8 \\
\hline $f_0^\prime(1500)$
& 1565.9$\pm$16.5 & 646.4$\pm$42.0  & 1563.2$\pm$16.8 & 656.0$\pm$40.4 \\
\hline $f_0 (1710)$
& 1718.6$\pm$35.6 & 145.8$\pm$32.4  & 1724.5$\pm$36.7 & 184.6$\pm$40.6 \\
\hline
\end{tabular}}
\end{center}

\caption{Masses and total widths of the resonances, obtained in the
analysis of set II.} \label{tab:mass_width_II}
\begin{center}
{
\vskip-0.2truecm
\def\arraystretch{1.2}
\begin{tabular}{|c|c|c|c|c|} \hline {} & \multicolumn{2}{|c|}{A-solution}&
\multicolumn{2}{|c|}{B-solution} \\\hline {State} & $m_{res}$[MeV] &
$\Gamma_{tot}$[MeV] & $m_{res}$[MeV] & $\Gamma_{tot}$[MeV]
\\ \hline $f_0 (600)$
&  652.3$\pm$11.8 & 828.6$\pm$14.0  &  743.4$\pm$14.1 &1001.8$\pm$17.4 \\
\hline $f_0 (980)$
& 1004.4$\pm$1.8  &  54.4$\pm$3.6   & 1004.0$\pm$1.9  &  56.4$\pm$3.8 \\
\hline $f_0 (1370)$
& 1367.8$\pm$13.5 & 444.8$\pm$53.8  & 1350.1$\pm$38.0 & 491.4$\pm$121.6 \\
\hline $f_0(1500)$
& 1498.7$\pm$5.3  & 113.6$\pm$9.8   & 1499.4$\pm$5.3  & 115.2$\pm$9.8 \\
\hline $f_0^\prime(1500)$
& 1563.8$\pm$30.1 & 656.8$\pm$75.6  & 1564.1$\pm$29.7 & 658.8$\pm$68.4 \\
\hline $f_0 (1710)$
& 1721.1$\pm$73.2 & 223.8$\pm$152.4 & 1721.3$\pm$47.1 & 223.8$\pm$105.0 \\
\hline
\end{tabular}}
\end{center}
\end{table}
Masses and total widths of the resonances, obtained in the analyses
of both sets of data (Tables~\ref{tab:mass_width_I}
and~\ref{tab:mass_width_II}), are reasonably close each other
taking into account their errors.

In Table \ref{tab:scattering_length} we compare our results for the
$\pi\pi$ scattering length $a_0^0$, obtained in the analyses of the
data of sets I and II, with results of some other theoretical and
experimental works.

%
%
\begin{table}[htb!]
\caption{The $\pi\pi$ scattering length $a_0^0$.}
\label{tab:scattering_length}
\begin{center}
{
\def\arraystretch{1.15}
\vskip-0.2truecm
\hspace*{-.5cm}
\begin{tabular}{|c|l|l|} \hline $a_0^0$
~[$m_{\pi^+}^{-1}$] & ~~~~~~~~~~Remarks & ~~~~~~~~~References
\\
\hline $0.222\pm 0.008$ & Analysis only of $\pi\pi$ scattering & This paper \\
\hline $0.230\pm 0.004$ & A-solution, set I
& This paper: \\
$0.282\pm 0.003$ & B-solution, set I
& combined analysis of \\
$0.226\pm 0.004$ & A-solution, set II &
processes $\pi\pi\to\pi\pi,K\overline{K}$\\
$0.275\pm 0.004$ & B-solution, set II &{}\\
\hline $0.26\pm 0.05$ & Analysis of the $K\to\pi\pi e\nu$ &
Ref.~\cite{Rosselet77} \\ {} & using Roy's equation &
{}
\\ \hline
$0.24\pm 0.09$ & Analysis of $\pi^-p\to\pi^+\pi^-n$ &
Ref.~\cite{Bel'kov79} \\ {} & using the effective range formula & {}
\\ \hline
$0.2220\pm0.0128_{\rm stat}$
& Experiment on $K_{e4}$
decay  & Ref.~\cite{Batley-epjc10} \\
$\pm0.0050_{\rm syst}\pm0.0037_{\rm th}$ &  &\\
\hline
$0.220 \pm 0.005$ & ChPT + Roy's equations &
Ref.~\cite{CGL00,Colangelo01} \\
\hline
$0.220\pm 0.008$ & Dispersion relations and $K_{e4}$ data &
Ref.~\cite{Garcia11} \\ \hline
$0.26$ & NJL model & Ref.~\cite{Volkov86} \\
\hline $0.28$ & NJL model & Ref.~\cite{Ivan_Troi95} \\
\hline
\end{tabular}}
\end{center}

%
%
\caption{The $\pi\pi$ scattering slope parameter $b_0^0$.}
\label{slope}
\begin{center}
\def\arraystretch{1.15}
\vskip-0.2truecm
\begin{tabular}{|c|l|l|}
\hline
$b_0^0$ ~[$m_{\pi^+}^{-3}$] & ~~~~~~~Remarks & ~~~~~~~~~~~~~~~~~References \\
\hline
$0.295\pm 0.021$ & Analysis only of $\pi\pi$ scattering & This paper \\
\hline
$0.210\pm 0.010$ & A-solution, set I & This paper: \\
$0.201\pm 0.007$ & B-solution, set I & combined analysis \\
$0.209\pm 0.011$ & A-solution, set II &
of processes $\pi\pi\to\pi\pi,K\overline{K}$\\
$0.208\pm 0.011$ & B-solution, set II &{}\\
\hline
$0.278\pm 0.005$ & Analysis using Roy-like equations &
Ref.~\cite{Garcia11} \\
{} & and forward dispersion relations & {} \\ \hline
$0.290\pm 0.006$ & Analysis using Roy's equations &
Ref.~\cite{KaminskiPRD77} \\
{} & and forward dispersion relations & {}
\\ \hline
\end{tabular}
\end{center}
\end{table}

Let us emphasize that in the case when we limit ourselves to the analysis
of the $\pi\pi$ scattering and in the case of the scenario A (lower mass
and width of $f_0(600)$ meson) we reproduce with a high accuracy the results
of the chiral perturbation theory (ChPT)~\cite{Colangelo01,CGL00} including
constraints imposed by the Roy's equations.
On the other side, the solutions of the scenario B (with a heavier mass and
width of $f_0(600)$ meson) is similar to the predictions of the chiral
approaches based on the linear realization of chiral symmetry (models
of the Nambu-Jona-Lasinio (NJL) type~\cite{Volkov86,Ivan_Troi95})
Taking into account very precise experiments at CERN performed 
by the NA48/2 Collaboration~\cite{Batley-epjc10} and  
the DIRAC Collaboration~\cite{DIRAC2011},
which confirmed the prediction of the ChPT~\cite{Colangelo01,CGL00} 
we should prefer the A-solution. In particular, 
the NA48/2 Collaboration~\cite{Batley-epjc10} 
extracted the $S$-wave $\pi\pi$ scattering lengths 
\eq
a_0^0&=&(0.2220 \pm 0.0128_{\rm stat} \pm 0.0050_{\rm syst} \pm
0.0037_{\rm th}) \, m_{\pi^+}^{-1}\,, \nonumber\\
a_2^0&=&(-0.0432\pm0.0086_{\rm stat}\pm0.0034_{\rm syst}\pm0.0028_{\rm th})
\, m_{\pi^+}^{-1}
\en
from the analysis of the $K_{e4}$ decay $K^\pm \to \pi^+\pi^- e^\pm \nu$.
The DIRAC Collaboration extracted the quantity
\eq
|a_0^0-a_2^0|=\left({0.2533^{+0.0080}_{-0.0078}}\biggl|_{\rm stat}
{}^{+0.0078}_{-0.0073} \biggl|_{\rm syst}\right)m_{\pi^+}^{-1}
\en
from the measurement of the $\pi^+\pi^-$ atom lifetime $\tau$ in the
ground state using the model-independent formula derived
in Refs.~\cite{HadAt} at next-to-leading order (NLO) in isospin breaking:
\eq
\tau^{-1} \sim (a_0^0 - a_2^0)^2 \, (1 + \delta) \,,
\en
where the quantity $\delta = (5.8 \pm 1.2) \times 10^{-2}$
encodes the NLO isospin-breaking correction.

In Table~\ref{slope} we show results for the slope parameter $b_0^0$
defined in following expansion around the threshold~\cite{Garcia11}:
\eq
\frac{\sqrt{s}}{4\,m_{\pi^+}} \sin 2\delta_{11}(s) = a_0^0\,k + b_0^0\,k^3
+ {\cal O}(k^5)\,,
\en
where $k= \sqrt{s/4-m_{\pi^+}^2}$ is the pion c.m. momentum. Our results agree
well with the other results only in the case of the one-channel analysis (only
$\pi\pi$ scattering).
In the combined analysis the obtained values are by 20-30\% smaller than
the results consistent with the dispersion relations.
The results do not differ so much for the A and B solutions as the scattering
length.

For convenience of usage of our results, we show the positions of zeros
on the $v$-plane, which correspond to the resonances, obtained for
the solutions A: \\
for set I --
\eq
{\rm for}~f_0(600):
        & &v_1= 1.3600\pm 0.0080 + (0.3797\pm 0.0076)i\,, \nonumber\\
        & &v_2= 0.6660\pm 0.0130 - (0.3254\pm 0.0079)i\,, \nonumber\\
{\rm for}~f_0(980):
        & &v_1^\prime= 1.0657\pm 0.0031 + (0.0346\pm 0.0023)i\,,
\nonumber\\
        & &v_2^\prime= 0.9006\pm 0.0019 - (0.0725\pm 0.0034)i\,,
\nonumber\\
{\rm for}~f_0(1370):
        & &v_3=-1.2331\pm 0.0010 - (0.0419\pm 0.0023)i\,, \nonumber\\
        & &v_4= 0.7966\pm 0.0011 - (0.0374\pm 0.0020)i\,, \nonumber\\
{\rm for}~f_0(1500):
        & &v_3^\prime=-1.2467\pm 0.0007 - (0.0077\pm 0.0006)i\,,
\nonumber\\
        & &v_4^\prime= 0.8023\pm 0.0004 - (0.0050\pm 0.0003)i\,, \\
{\rm for}~f_0^\prime(1500):
        & &v_5= 1.2641\pm 0.0014 + (0.0362\pm 0.0016)i\,, \nonumber\\
        & &v_6=-1.2643\pm 0.0032 - (0.0386\pm 0.0045)i\,, \nonumber\\
        & &v_7= 0.7981\pm 0.0015 - (0.0184\pm 0.0012)i\,, \nonumber\\
        & &v_8= 0.7877\pm 0.0033 - (0.0265\pm 0.0031)i\,,
\nonumber\\
{\rm for}~f_0(1710):
        & &v_3^{\prime\prime}=-1.2701\pm 0.0030 - (0.0062\pm 0.0011)i\,,
        \nonumber\\
        & &v_4^{\prime\prime}= 0.7881\pm 0.0018 - (0.0042\pm 0.0007)i\,;
        \nonumber
\en
for set II --
\eq
{\rm for}~f_0(600):
       & &v_1= 1.3815\pm 0.0102 + (0.3713\pm 0.0096)i\,, \nonumber\\
       & &v_2= 0.6360\pm 0.0159 - (0.3286\pm 0.0089)i\,, \nonumber\\
{\rm for}~f_0(980):
       & &v_1^\prime= 1.0664\pm 0.0022 + (0.0373\pm 0.0021)i\,,
\nonumber\\
       & &v_2^\prime= 0.8998\pm 0.0017 - (0.0742\pm 0.0025)i\,,
\nonumber\\
{\rm for}~f_0(1370):
       & &v_3=-1.2342\pm 0.0013 - (0.0413\pm 0.0027)i\,, \nonumber\\
       & &v_4= 0.7970\pm 0.0012 - (0.0393\pm 0.0025)i\,, \nonumber\\
{\rm for}~f_0(1500):
       & &v_3^\prime= 1.2463\pm 0.0006 - (0.0078\pm 0.0006)i\,,
\nonumber\\
       & &v_4^\prime= 0.8029\pm 0.0003 - (0.0054\pm 0.0003)i\,,
\\
{\rm for}~f_0^\prime(1500):
       & &v_5= 1.2642\pm 0.0059 + (0.0369\pm 0.0103)i\,, \nonumber\\
       & &v_6=-1.2640\pm 0.0248 - (0.0389\pm 0.0384)i\,, \nonumber\\
       & &v_7= 0.7980\pm 0.0009 - (0.0185\pm 0.0010)i\,, \nonumber\\
       & &v_8= 0.7880\pm 0.0128 - (0.0268\pm 0.0202)i\,, \nonumber\\
{\rm for}~f_0(1710):
       & &v_3^{\prime\prime}=-1.2708\pm 0.0053 - (0.0095\pm 0.0034)i\,,
       \nonumber\\
       & &v_4^{\prime\prime}= 0.7884\pm 0.0031 - (0.0041\pm 0.0010)i\,.
       \nonumber
\en

\section{Conclusions}

One of the objectives of this paper was to demonstrate that 
the parameters of wide multichannel resonances can not be determined 
in principle by the dispersion equation approach independent on accuracy 
of data. Therefore, we had to return back to two-channel consideration 
because the two-channel approach is free of any assumptions unlike 
three-channel approach, where we had to built 4-sheet model of initial 
8-sheet Riemann surface. 
Obtained parameters in two-channel approach are differed from 
results of one-channel approach. However their values are quite close 
to the parameters of three-channel analysis. In three-channel analysis 
we can not obtain the $\pi\pi$ scattering lenghts, while two-channel 
approach suits for determining the low-energy parameters of $\pi\pi$ 
scattering. As byproduct of two-channel consideration 
we obtained two solutions for the $a_0^0$ scattering length: the solution A 
corresponds to the chiral perturbation theory based on nonlinear realization 
of chiral symmetry and standard scenario for quark condensate, while 
the solution B corresponds to the linear realization of chiral symmetry. 

Before discussing the results of performed analysis, let us note that
for calculating the wide resonance parameters (masses, total widths
and coupling constants with channels) it is vital to use the poles on
those sheets on which the poles are not shifted due to a coupling of
channels because these poles respect positions of zeros on the physical
sheet. These appropriate sheets are numerated by $2^i$ ($i=1,\cdots,N$
is the number of channel), i.e. II, IV, VIII,$\cdots$. This conclusion
is model-independent. In this work we demonstrated this principle on
the basis of analytic continuations of the $S$-matrix elements to
unphysical sheets in the two-channel case. The general case of N channels
can be found in other our papers~\cite{SBL-arXiv11,KMS96}.

It appears that neglecting the above-indicated principle can cause
misunderstandings. This concerns especially the analyses which do not
consider the structure of the Riemann surface of the $S$-matrix.
For example, in literature there is a common opinion (delusion) that
the resonance parameters should be calculated using resonance poles
nearest to the physical region. This is right only in the one-channel
case. In the multi-channel case this is not correct.
It is obvious that, e.g., the resonance pole on sheet~III, which is
situated above the second threshold, is nearer to the physical region
than the pole on sheet~II from the pole cluster of the same resonance
because above the second ($K\overline{K}$) threshold the physical
region (an upper edge of the right-hand cut) is joined directly with
sheet~III. Therefore, the pole on sheet~III influences most strongly
on the energy behavior of the amplitude and this pole will be found
in the analyses, not taking into account the structure of the Riemann
surface and the representation of resonances by the pole clusters.

In our model-independent approach using the uniformizing variable, we
analyzed data on isoscalar S-wave processes $\pi\pi\to\pi\pi,K\overline{K}$
including the very precise NA48/2-Collaboration $\pi\pi$-data in the
threshold region. Moreover, for the $\pi\pi$ scattering the alternative
data were taken: these are the data by B.~Hyams et al.(1973) (set~I)
and by R.~Kami\'nski et al.(2002) (set~II) which are considerably
different in energy regions 1.55-1.85 and 1.25-1.55~GeV.

When analyzing only the $\pi\pi$ scattering data from set~I, it was shown
that a good description from the $\pi\pi$ threshold up to 1.89~GeV is
achieved ($\chi^2/\mbox{NDF}\approx1.07$) with parameters of resonances
(Table \ref{tab:pole_clusters_pipi}) mainly coinciding with the ones
cited as estimation of the PDG \cite{PDG} where for the $f_0(600)$
the found pole on sheet II coincides practically with the one at around
$450 - i275$~MeV which was found in the recent dispersive
$\pi\pi$-scattering data analyses \cite{Garcia11,Colangelo01} where
the amplitude is taken on the 2-sheeted Riemann surface -- unlike our
4-sheeted consideration -- and the good description is obtained only to
about 1.15~GeV.
The only parameter strongly differing from the ones, cited by PDG as
preferred, is the width of $f_0(1500)$ (336~MeV against 109$\pm$7~MeV
of PDG). As to the $f_0(1500)$, we analyzed also the $\pi\pi$-scattering
assuming two states (narrow and wide) in this region. 
Note, appearing of two resonances instead of one is just result of using 
several channel model. 
The less channels lead to less poles and their interference 
what reduces possibility of their interpretation.
Description is of the same accuracy as in the first case. 
Parameters of the narrow
$f_0(1500)$ coincide with those preferred by the PDG. I.e., when
analyzing the $\pi\pi$ scattering data from set~I, there is admitted
the above two-resonance interpretation in the 1500-MeV region, because
the narrow state, described by the pole cluster of type~({\bf b}),
does not influence strongly the $\pi\pi$-scattering phase-shift
behavior. On the contrary, when analyzing the $\pi\pi$ scattering data
from set~II, it is necessary to use the two-resonance description in
the 1500-MeV region with the representation of the narrow state by the
pole cluster of type~({\bf a}) that provides more rapid growth of the
$\pi\pi$ phase shift above 1.4~GeV than in the set~I. 

However, first, the satisfactory description still does not mean that
it is the adequate description. The point is that the negative
phase-shift in the background arises already on the $\pi\pi$ threshold.
This is denoted as a pseudo-background. It appears to compensate for a
too fast rise of the phase-shift of the amplitude, which is induced by
the parameters of the $f_0(600)$, i.e., it indicates that these
parameters are incorrect. Especially, the non zero negative phase-shift
is in contradiction with the expectation that in our parametrization
the phase shift in the $\pi\pi$ background below the $K\overline{K}$
threshold is practically zero \cite{SKN-epja02} because the left-hand
branch-point at $s=0$, which gives a main contribution to the $\pi\pi$
background below the $K\overline{K}$ threshold, is included explicitly
in the uniformizing variable.
Other possible contributions of the left-hand cut from exchanges of
the lightest mesons -- the $\rho$-meson and the $f_0 (600)$ --
practically obliterate each other because vector and scalar particles
contribute with the opposite signs due to gauge invariance.

Second, a description of the process $\pi\pi\to K\overline{K}$ with
the resonance parameters obtained in the analysis of only the $\pi\pi$
scattering is satisfactory only for the phase shift which is due to
the fact that we approximate the left-hand branch-point at
$s=4(m_K^2-m_\pi^2)$ in $S_{12}$ and $S_{22}$ by the pole of the
fourth power and that the pole clusters of resonances are chosen
correctly. The module of the $S$-matrix element is described
satisfactorily only from the $K\overline{K}$ threshold up to the
energy about 1.15~GeV as it should be due to the two-channel unitarity
(see eqs. (\ref{eq:2-ch-uni})). Above this value of energy the
module is not described well even qualitatively.

To this point, let us also note results of our previous
work~\cite{SBL-arXiv11} for the coupling constants of the ${f_0}$
mesons with various channels. Despite a preliminary character
of these results, one can draw some conclusions about, e.g., the
$f_0(600)$ and $f_0(980)$. These states turn out to have large
coupling constants with the $K\overline{K}$ and especially
$\eta\eta$ systems, i.e., studying these states we deal with
a multi-channel problem. Even if these states can not decay into the
$\eta\eta$ channel, their large coupling with the $\eta\eta$ system
should manifest itself in exchanges in the $\pi\eta$ scattering.

It was shown in the two-channel approach to the $\pi\pi$ scattering,
that the combined analysis of the coupled processes -- the $\pi\pi$
scattering and $\pi\pi\to K\overline{K}$ -- is needed. This analysis
was done for two sets of data (sets I and II) including two resonances
in the 1500-MeV region. Then both above-indicated important remarks,
related to the analysis only of the $\pi\pi$-scattering, are ruled out.
In the combined analysis the parameters of the $f_0(600)$ were changed
considerably with the new values closer to those obtained in our
previous three-channel analysis \cite{SBL-arXiv11}. It was shown for
the data of sets I and II, that in the region below 1~GeV, there are
two solutions, A and B, related to the $\sigma$-meson/$f_0 (600)$
with the mass about 0.65~GeV and width about 0.8~GeV in the case A
and $m_\sigma\approx m_\rho$ and width about 1~GeV in the case B.
This agrees with the Weinberg prediction done on the basis of a
mended symmetry~\cite{Wei90}. Moreover, this is also in agreement
with a refined analysis using the large-$N_c$ consistency conditions
between the unitarization and resonance saturation suggesting
$m_\rho-m_\sigma=O(N_c^{-1})$ \cite{Nieves-Arriola}.
Note the prediction of a soft-wall AdS/QCD approach~\cite{GLSV_13}
for the mass of the lowest $f_0$ meson -- 721~MeV -- is in some
agreement with our result in the A solutions and practically coincides
with the one in B solutions. 

Following a tradition, we speak here on the masses and total widths
of resonances though the broad multi-channel states are represented
more correctly by the pole clusters which are their model-independent
characteristics (see the discussion in Sec.III) whereas the masses and
widths are very model-dependent for wide resonances. Values of masses
are necessary, e.g. for the mass relations of multiplets.

The obtained values for the $\pi\pi$ scattering length for the
A-solutions are in accordance with predictions of ChPT (non-linear
realization of chiral symmetry), whereas the values for the
B-solutions agree with predictions of chiral theory with linear
realization of chiral symmetry (models of NJL type).
Generally, considering only description of the analyzed processes,
it is impossible for now to prefer any of these solutions.
The B-solutions for set~I and II describe the data slightly better,
whereas the obtained $\pi\pi$ scattering lengths for the A-solutions
have more acceptable values. However, if one considers the problem of
precise determination of the $\pi\pi$ scattering length $a_0^0$ to be
solved taking into account the results of the NA48/2
Collaboration~\cite{Batley-epjc10} and the DIRAC
experiment~\cite{DIRAC2011} at CERN, then the A-solutions should be
chosen. Therefore, our final conclusion is that the agreement of our
approach with ChPT and data for the $a_0^0$ $\pi\pi$ scattering length
favors to the A-solutions for the masses and widths of the scalar
resonances.

Finally, let us stress that our method is developed under the first
principles such as analyticity, unitarity and Lorentz invariance,
and therefore, it is free of any suppositions on dynamics except for
an obvious statement that a main model-independent contribution of resonances
is given by the pole clusters and possible remaining small (model-dependent)
contributions of resonances can be included in the background.

\section*{Acknowledgments}

The authors thank Thomas Gutsche and Mikhail Ivanov for useful
discussions and interest in this work.
This work was supported in part by the Grant Program of Plenipotentiary
of Slovak Republic at JINR, the Heisenberg-Landau Program, the
Votruba-Blokhintsev Program for Cooperation of Czech Republic with JINR,
the Grant Agency of the Czech Republic (grant No. P203/12/2126),
the Bogoliubov-Infeld Program for Cooperation of Poland with JINR,
the DFG under Contract No. LY 114/2-1, and the Polish Ministry of Science
and Higher Education (grant No N N202 101 368).
The work was also partially supported under the project 2.3684.2011 of
Tomsk State University.

\end{document}